# The Fluctuation Theorem
# and Green-Kubo Relations

Debra J. Searles* and Denis J. Evans#

*Department of Chemistry, University of Queensland,

Brisbane, QLD 4072, Australia

#Research School of Chemistry, Australian National University, GPO Box 414,

Canberra, ACT 2601, Australia

**Abstract**

Green-Kubo and Einstein expressions for the transport coefficients of a fluid in a nonequilibrium steady state can be derived using the Fluctuation Theorem and by assuming the probability distribution of the time-averaged dissipative flux is Gaussian. These expressions are consistent with those obtained using linear response theory and are valid in the linear regime. It is shown that these expressions are however, not valid in the nonlinear regime where the fluid is driven far from equilibrium. We advance an argument for why these expressions are only valid in the linear response, zero field limit.

## I. INTRODUCTION

In 1993 Evans, Cohen and Morriss [1], ECM2, gave a quite general formula for the logarithm of the probability ratio that in a nonequilibrium steady state, the time averaged dissipative flux takes on a value, $\bar{J}_+(t)$, to minus that value, namely, $\bar{J}_-(t) = -\bar{J}_+(t)$. That is they gave a formula for $\ln[p(\bar{J}_+(t)) / p(\bar{J}_-(t))]$ from a natural invariant measure [1, 2]. This formula gives an analytic expression for the probability that, for a finite system and for a finite time, the dissipative flux flows in the reverse direction to that required by the Second Law of Thermodynamics. The formula has come to be known as the Fluctuation Theorem, FT. Surprisingly perhaps, it is valid far from equilibrium in the nonlinear response regime [1]. Since 1993 there have been a number of derivations and generalisations of the FT. Evans and Searles [3-5] gave a derivation, similar to that given here, which considered transient, rather than steady state, nonequilibrium averages and employed the Liouville measure. Gallavotti and Cohen [6,7], gave a proof of the formula for a nonequilibrium stationary state, based on a Chaotic Hypothesis and employing the SRB measure. In the long time limit, when steady state averages are independent of the initial phase used to generate the steady state trajectory, averages over *transient* segments which originate from the initial equilibrium microcanonical ensemble can be expected to approach those taken over nonequilibrium *steady state* segments. Thus for chaotic systems both approaches should be able to explain the steady state results. However this point is being debated [8]. Other generalisations of the FT have recently been developed [9-12].

In a footnote to their original paper ECM2 also pointed out that in the weak field regime, there was a connection between the FT, the Central Limit Theorem (CLT) [13, 14], and Green-Kubo relations [1, 4]. In the present paper we explore this connection further and consider the validity of the Green-Kubo relations far from equilibrium. We show that ***if*** the distribution of the time averaged dissipative flux, $p(\bar{J}(t))$, is Gaussian arbitrarily far from the mean, then from the FT one can derive both generalised Einstein and generalised Green-Kubo relations for the relevant transport coefficient. Both isothermal (ie isokinetic) and isoenergetic dynamics are considered. We conduct computer simulations which prove that outside the linear regime, these generalised Green-Kubo and Einstein relations are incorrect.

It turns out, that in order for the nonlinear Green-Kubo relations to be valid $p(\bar{J}(t)/\sigma_{\bar{J}(t)})$ must be a normalised Gaussian when *both* t *and* $\bar{J}(t)/\sigma_{\bar{J}(t)} \to \infty$. However, this is not guaranteed by the Central Limit theorem [15] and the nonlinear Green-Kubo relations are invalid.

## II. NEMD DYNAMICAL SYSTEMS

The development of NonEquilibrium Molecular Dynamics, NEMD, over the previous two decades has lead to a set of deterministic algorithms (ie N-particle dynamical systems) from which one can in principle, calculate correct values for each of the Navier-Stokes transport coefficients [16]. These dynamical systems actually duplicate the salient features of real experimental nonequilibrium steady states. In the linear regime close to equilibrium, nonequilibrium statistical mechanics is used to prove that in the large system limit the calculated transport properties are correct. Using NEMD one can calculate far more than just transport coefficients. One can also correctly calculate the changes to the local molecular structure and dynamics, caused by the applied external fields.

Consider an N-particle system in 3 Cartesian dimensions, with coordinates and peculiar momenta, $\{\mathbf{q}_1,\mathbf{q}_2,..\mathbf{q}_N,\mathbf{p}_1,..\mathbf{p}_N\} \equiv (\mathbf{q},\mathbf{p}) \equiv \boldsymbol{\Gamma}$. The internal energy of the system is $H_0 \equiv \sum_{i=1}^{N} p_i^2 / 2m + \Phi(\mathbf{q})$ where $\Phi(\mathbf{q})$ is the interparticle potential energy which is a function of the coordinates of all of the particles, $\mathbf{q}$. In the presence of an external field $F_e$, the thermostatted equations of motion are taken to be,

$$\dot{\mathbf{q}}_i = \mathbf{p}_i / m + \mathbf{C}_i(\boldsymbol{\Gamma})F_e$$

$$\dot{\mathbf{p}}_i = \mathbf{F}_i(\mathbf{q}) + \mathbf{D}_i(\boldsymbol{\Gamma})F_e - \alpha(\boldsymbol{\Gamma})\mathbf{p}_i \qquad (1)$$

where

$$\mathbf{F}_i(\mathbf{q}) = -\partial\Phi(\mathbf{q}) / \partial\mathbf{q}_i \qquad (2)$$

and $\alpha$ is the thermostat multiplier derived from Gauss' Principle of Least Constraint in order to fix the peculiar kinetic energy, $K \equiv \sum_{i=1}^{N} p_i^2 / 2m$, or the internal energy, $H_0$. In a constant energy system the thermostat multiplier is easily seen to be,

$$\alpha_E = -J(\boldsymbol{\Gamma})VF_e / 2K \qquad (3)$$

while in an isokinetic system the corresponding expression for the multiplier is,

$$\alpha_K = \sum \frac{\mathbf{p}_i}{m} \bullet (\mathbf{F}_i + \mathbf{D}_i F_e) \Big/ 2K \,. \tag{4}$$

We note that the thermostatted equations of motion are time reversible. The dissipative flux is defined in terms of the adiabatic (ie unthermostatted) derivative of the internal energy,

$$\dot{H}_0^{ad} \equiv -J(\mathbf{\Gamma})VF_e \tag{5}$$

where V is the system volume.

In an isokinetic system, the balance between the work done on the system by the external field and the heat removed by the thermostat implies that

$$\lim_{(t\to\infty)} \int_0^t ds\, J(\mathbf{\Gamma}(s))VF_e = -2K_0 \lim_{(t\to\infty)} \int_0^t ds\, \alpha_K(\mathbf{\Gamma}(s)), \tag{6}$$

while in a constant energy system, energy balance is exact instantaneously,

$$J(\mathbf{\Gamma})VF_e = -2K(\mathbf{p})\alpha_E(\mathbf{\Gamma}). \tag{7}$$

In equation (6), $K_0$ is the (fixed) peculiar kinetic energy,

$$3Nk_BT/2 \equiv 3N\beta_0^{-1}/2 \equiv K_0 \,. \tag{8}$$

A shorthand notation will be used to refer to the time averaged value of a phase function along a trajectory segment, $\mathbf{\Gamma}_+(s);\, 0 < s < t$. We will write,

$$\overline{A}_+(t) \equiv \tfrac{1}{t} \int_0^t ds\, A(\mathbf{\Gamma}_+(s)). \tag{9}$$

Since the dynamics is time reversible, for every trajectory segment $\mathbf{\Gamma}_+(s);\, 0 < s < t$, there exists an antisegment, $\mathbf{\Gamma}_-(s);\, 0 < s < t$. A plus or minus sign is ascribed to a particular trajectory segment depending on the sign of the time averaged value of the thermostat multiplier: therefore by definition $\overline{\alpha}_+(t) > 0$. The time reversed conjugate of the segment $\mathbf{\Gamma}_+(s);\, 0 < s < t$, namely $\mathbf{\Gamma}_-(s);\, 0 < s < t$, is termed an antisegment and,

$$\overline{A}_{-}(t) \equiv \frac{1}{t}\int_0^t ds\, A(\mathbf{\Gamma}_{-}(s)). \tag{10}$$

Depending on the parity of the phase function $A(\mathbf{\Gamma})$ under the time reversal mapping there may be a simple relation between $\overline{A}_{+}(t)$ and $\overline{A}_{-}(t)$. Without loss of generality we take the external field to be even under time reversal symmetry, therefore the dissipative flux is odd and,

$$\overline{J}_{-}(t) = -\overline{J}_{+}(t), \forall t. \tag{11}$$

Using this notation the dissipative flux is related to the phase space compression accomplished by the thermostat,

$$\begin{aligned} &\lim_{(t\to\infty)} \beta\overline{J}_{+}(t)VF_e = -\lim_{(t\to\infty)} 3N\overline{\alpha}_{+}(t) && \text{isokinetic} \\ &\overline{\beta J}_{+}(t)VF_e = -3N\overline{\alpha}_{+}(t) && \text{isoenergetic} \end{aligned} \tag{12}$$

where for both the isokinetic and isoenergetic systems,

$$\beta J(\mathbf{\Gamma})V \equiv 3NJ(\mathbf{\Gamma})V / 2K(\mathbf{p})$$

but in the isokinetic case, the peculiar kinetic energy K is a constant of the motion. Since $\beta$ is always positive, we see from (12), that the sign convention for distinguishing segments and antisegments can equally well be taken from the sign of the dissipative flux.

## III. THE (TRANSIENT) FLUCTUATION THEOREM

For our system, since the adiabatic incompressibility of phase space (AI$\boldsymbol{\Gamma}$) holds [16], the Liouville equation for the N-particle distribution function $f(\boldsymbol{\Gamma},t)$, reads,

$$\frac{df(\boldsymbol{\Gamma},t)}{dt} = -f(\boldsymbol{\Gamma},t)\partial\dot{\boldsymbol{\Gamma}}\bullet/\partial\boldsymbol{\Gamma} = 3N\alpha(\boldsymbol{\Gamma})f(\boldsymbol{\Gamma},t) + O(1)f(\boldsymbol{\Gamma},t). \quad (13)$$

The O(1) terms are omitted in the following discussion. Incorporation of these terms poses no difficulty but complicates the expressions and the consequences can be neglected in the large system limit. The solution of this equation can be written as [4]

$$\begin{aligned} f(\boldsymbol{\Gamma}_\pm(t),t) &= \exp[\int_0^t 3N\alpha(\boldsymbol{\Gamma}_\pm(s))ds]f(\boldsymbol{\Gamma}_\pm,0) \\ &= \exp[3N\overline{\alpha}_\pm(t)t]f(\boldsymbol{\Gamma}_\pm,0) \end{aligned} \quad . \quad (14)$$

This is known as the Lagrangian form of the Kawasaki distribution function [4].

Consider the propagation of a phase point along a trajectory in phase space. If we select an initial, t = 0, phase, $\boldsymbol{\Gamma}_{(1)}$, and we advance time from 0 to $\tau$ using the equations of motion (1) we obtain $\boldsymbol{\Gamma}_{(2)} = \boldsymbol{\Gamma}(\tau;\boldsymbol{\Gamma}_{(1)}) = \exp[iL(\boldsymbol{\Gamma}_{(1)},F_e)\tau]\boldsymbol{\Gamma}_{(1)}$, where the phase Liouvillean, $iL(\boldsymbol{\Gamma}_{(1)},F_e)$, is defined as, $iL(\boldsymbol{\Gamma},F_e) \equiv \dot{\mathbf{q}}(\boldsymbol{\Gamma},F_e)\bullet\partial/\partial\mathbf{q} + \dot{\mathbf{p}}(\boldsymbol{\Gamma},F_e)\bullet\partial/\partial\mathbf{p}$. Continuing on to $2\tau$ gives $\boldsymbol{\Gamma}_{(3)} = \exp[iL(\boldsymbol{\Gamma}_{(2)},F_e)\tau]\boldsymbol{\Gamma}_{(2)} = \exp[iL(\boldsymbol{\Gamma}_{(1)},F_e)2\tau]\boldsymbol{\Gamma}_{(1)}$. This is demonstrated in Figure 1.

From this trajectory segment, we can construct a time-reversed trajectory. At the *midpoint* of the trajectory segment $\boldsymbol{\Gamma}_{(1,3)}$ (i.e. at t = $\tau$) we apply the time reversal mapping, $M^{(T)}$, to $\boldsymbol{\Gamma}_{(2)}$ generating $M^{(T)}\boldsymbol{\Gamma}_{(2)} \equiv \boldsymbol{\Gamma}_{(5)}$. If we now propagate backward in time keeping the same external field, we obtain $\boldsymbol{\Gamma}_{(4)} = \exp[-iL(\boldsymbol{\Gamma}_{(5)},F_e)\tau]\boldsymbol{\Gamma}_{(5)}$. $\boldsymbol{\Gamma}_{(4)}$ is the initial t = 0 phase from which a segment $\boldsymbol{\Gamma}_{(4,6)}$ can be generated with $\boldsymbol{\Gamma}_{(6)} = \exp[iL(\boldsymbol{\Gamma}_{(4)},F_e)2\tau]\boldsymbol{\Gamma}_{(4)}$. We denote the trajectory $\tau$-segment $\boldsymbol{\Gamma}_{(1)}\rightarrow\boldsymbol{\Gamma}_{(3)}$, as $\boldsymbol{\Gamma}_{(1,3)} = \boldsymbol{\Gamma}_+$; similarly $\boldsymbol{\Gamma}_{(4,6)} = \boldsymbol{\Gamma}_-$. Using the symmetry of the equations of motion it is trivial to show that, $J(\boldsymbol{\Gamma}_{(2)}) = -J(\boldsymbol{\Gamma}_{(5)})$ and that $J(t;\ \boldsymbol{\Gamma}_+,\ 0 < t < 2\tau) = -J(2\tau\text{-}t;\ \boldsymbol{\Gamma}_-,\ 0 < t < 2\tau)$, - see Figure 1. We now have an algorithm for finding *initial* phases which will subsequently generate the conjugate segments.

The ratio of probabilities of finding the initial phases $\boldsymbol{\Gamma}_{(1)}, \boldsymbol{\Gamma}_{(4)}$ which generate these conjugate segments will now be discussed. In a *causal* universe, the probabilities of observing the segments $\boldsymbol{\Gamma}_{+}$ and $\boldsymbol{\Gamma}_{-}$ are proportional to the probabilities of observing the *initial* phases which generate those segments [3, 5]. It is convenient to consider a small phase space volume, $\delta V(\boldsymbol{\Gamma}_{(i)}(0))$ about an initial phase, $\boldsymbol{\Gamma}_{(i)}(0)$. If we are considering isoenergetic dynamics, then the initial equilibrium phases are distributed *microcanonically*, and therefore the probability of observing ensemble members inside $\delta V(\boldsymbol{\Gamma}_{(i)}(0))$, is proportional to $\delta V(\boldsymbol{\Gamma}_{(i)}(0))$ (for generalisations to other ensembles see [17]). From the Liouville equation (13) and the fact that for sufficiently small volumes, $\delta V(\boldsymbol{\Gamma}(t)) \sim 1/f(\boldsymbol{\Gamma}(t),t)$, we can make the following observations: $\delta V_2 = \delta V_1(\tau) = \delta V_1(0)\exp[-\int_0^{\tau} 3N\alpha(s;\Gamma_{(1)})ds]$ and, $\delta V_3=\delta V_1(2\tau)= \delta V_1(0)\exp[-\int_0^{2\tau} 3N\alpha(s;\boldsymbol{\Gamma}_{(1)})ds]$. Because the segment $\boldsymbol{\Gamma}_{(4,6)}$ is related to $\boldsymbol{\Gamma}_{(1,3)}$ by $M^T$ which is applied at $t = \tau$, and the Jacobian of $M^T$ is unity, $\delta V_2 = \delta V_5 \Rightarrow \delta V_3 = \delta V_4$ and $\delta V_1(0) = \delta V_6$.

However, since $\delta V_1(0)$ and $\delta V_4(0)$ are volumes at $t = 0$ and since the distribution of initial phases is microcanonical, we can compute the ratio of probabilities of observing $t = 0$ phases within $\delta V_1(0)$ and $\delta V_4(0)$. This ratio is just the volume ratio,

$$\delta V_4(0)/\delta V_1(0) = \delta V_1(2\tau)/\delta V_1(0) = \exp[\int_0^{2\tau} -3N\alpha(s;\boldsymbol{\Gamma}_{(1)})ds], \forall\tau. \quad (15)$$

There may be trajectory segments whose initial phases lie outside the phase space volume $\delta V_1(0)$ but which have the same value of $\overline{\alpha}(2t)$ as those lying inside $\delta V_1(0)$. Suppose that there are two noncontiguous subvolumes of phase space $\delta V_1(0)$, $\delta V_{1'}(0)$ from which trajectories originate which, after a time $2\tau$, have time averaged values of $\alpha$ which lie in the range: $(\overline{\alpha}_+(2\tau), \overline{\alpha}_+(2\tau) + d\alpha)$, (see Figure 2). Suppose that at time $2\tau$ these volumes evolve to: $\delta V_1(2\tau)=\delta V_3$, $\delta V_{1'}(2\tau)=\delta V_{3'}$. At time zero the corresponding volumes for the corresponding antitrajectories are: $\delta V_1(2\tau)=\delta V_4$, $\delta V_{1'}(2\tau)=\delta V_{4'}$.

The ratio of probabilities of observing trajectories $(\overline{\alpha}_+(2\tau), \overline{\alpha}_+(2\tau) + d\alpha)$ compared to the corresponding antitrajectories, $(\overline{\alpha}_-(2\tau), \overline{\alpha}_-(2\tau) + d\alpha)$, is

$$(\delta V_4 + \delta V_{4'}) \big/ (\delta V_1 + \delta V_{1'}) = (\delta V_3 + \delta V_{3'}) \big/ (\delta V_1 + \delta V_{1'})$$

$$\underset{\lim\ d\alpha\to 0}{=} (\delta V_1 \exp[3N\overline{\alpha}_+(2\tau)2\tau] + \delta V_{1'} \exp[3N\overline{\alpha}_+(2\tau)2\tau]) \big/ (\delta V_1 + \delta V_{1'})$$

$$= \exp[3N\overline{\alpha}_+(2\tau)2\tau] \tag{16}$$

since $\overline{\alpha}(2\tau)$ is the same for both trajectories. This shows that even when noncontiguous regions of phase space have the same time averaged values for the thermostatting multiplier, the ratio of probability that $\overline{\alpha}(t) = \overline{\alpha}_+(t) = A$ to the probability that $\overline{\alpha}(t) = \overline{\alpha}_-(t) = -A$, (ie $p(\overline{\alpha}(t) = A)/p(\overline{\alpha}(t) = -A)$, where A is any required value of the time average of $\alpha$ is,

$$\frac{p(\overline{\alpha}(t) = A)}{p(\overline{\alpha}(t) = -A)} = \exp[3NAt]. \tag{17}$$

This formula is exact for transient trajectory segments of a system undergoing isoenergetic dynamics. For isoenergetic dynamics there is a linear relationship between the value of $\alpha$ and the value of $\beta J$, so

$$\frac{p(\overline{\alpha}(t) = A)}{p(\overline{\alpha}(t) = -A)} = \frac{p(\overline{\beta J}(t) = B)}{p(\overline{\beta J}(t) = -B)} = \exp[3NAt] = \exp[-BVF_e t]. \tag{18}$$

where $B=-3NA/VF_e$. For isokinetic dynamics, the procedure above can be used to show that [17],

$$\frac{p(\overline{J}(t) = B)}{p(\overline{J}(t) = B)} = \exp[-B\beta VF_e t]. \tag{19}$$

If we are interested in steady state segments, equations (18) and (19) will only be true in the limit as the segment duration, t, goes to infinity [1, 6-8, 17] and only when the steady state is unique:

$$\lim_{(t\to\infty)} \frac{1}{t} \ln\left( \frac{p(\overline{\beta J}(t) = B)}{p(\overline{\beta J}(t) = B)} \right) = -BVF_e. \tag{20}$$

Equations (18, 19 and 20) express what has become known as the Fluctuation Theorem, (FT),

for the dissipative flux for isokinetic and the isoenergetic dynamics [1, 2-7, 17].

## IV. EINSTEIN AND GREEN-KUBO RELATIONS

We consider first the isokinetic case. In this case $\beta$ is a constant of the motion: $\overline{\beta J}(t) = \beta_0 \bar{J}(t)$. It might be expected that as the averaging time, t becomes arbitrarily large compared to the Maxwell time, $\tau_M$, which characterises serial correlations in the dissipative flux, contributions to the trajectory segment averages of the dissipative flux, $\{\bar{J}(t)\}$, would become statistically independent and therefore satisfy the Central Limit Theorem, (CLT). That is, as $t \to \infty$, the distribution would *approach* a Gaussian. If the distribution is Gaussian, it is trivial to show that there is a relation between the logarithm of conjugate probabilities of time averaged steady state dissipative fluxes and the variance of the distribution of those averaged dissipative fluxes,

$$\lim_{(t\to\infty)} \frac{1}{t} \ln\left( \frac{p(\overline{\beta J}(t) = \beta B)}{p(\overline{\beta J}(t) = -\beta B)} \right) = \lim_{(t\to\infty)} \frac{1}{t} \ln\left( \frac{p(\bar{J}(t) = B)}{p(\bar{J}(t) = B)} \right)$$
$$= \lim_{(t\to\infty)} \frac{2B\langle J\rangle_{F_e}}{t\sigma^2_{\bar{J}(t)}(t)} \tag{21}$$

where $\sigma^2_{\bar{J}(t)}(t)$ is the variance of the distribution of $\{\bar{J}(t)\}$. Combining this equation with (20) shows that if the distribution is Gaussian there must be a trivial relation between the variance and the mean of the distribution of averaged fluxes [18]. From this relation the nonlinear transport coefficient is given ,

$$L(F_e) = \frac{-\langle J\rangle_{F_e}}{F_e} = \lim_{(t\to\infty)} \tfrac{1}{2}\beta_0 V t \sigma^2_{\bar{J}(t)}. \tag{22}$$

In the zero field limit this equation constitutes an Einstein relation for the linear transport coefficient, L(0). Except for the case of colour conductivity where (22) is equivalent to the standard Einstein expression for the self diffusion coefficient [19], these zero field Einstein relations are not well known. For nonzero applied fields, the generalised Einstein relation for the field dependent transport coefficient, $L(F_e)$, (22) is, as we shall see, incorrect.

In the long time limit the variance of the steady state distribution of t-averaged fluxes,

$$\sigma^2_{\bar{J}(t)}(F_e) = \left\langle (\bar{J}(t) - \langle J \rangle_{F_e})^2 \right\rangle_{F_e} \quad , \tag{23}$$

satisfies the equation (see [4] and also the Appendix),

$$\begin{aligned} \lim_{(t\to\infty)} t\sigma^2_{\bar{J}(t)}(F_e) &= 2\tilde{L}_J(0;F_e) \big/ \beta_0 V + \lim_{(t\to\infty)} 2\tilde{L}'_J(0;F_e) \big/ \beta_0 V t \\ &= 2\tilde{L}_J(0;F_e) \big/ \beta_0 V \end{aligned} \tag{24}$$

where

$$\tilde{L}_J(s;F_e) \equiv \beta_0 V \int_0^\infty dt\ e^{-st} \left\langle (J(0) - \langle J \rangle_{F_e})(J(t) - \langle J \rangle_{F_e}) \right\rangle . \tag{25}$$

$\tilde{L}_J(s;F_e)$ is the frequency and field dependent Green-Kubo transform (GK), of the dissipative flux and therefore is essentially the Fourier-Laplace transform of the field dependent dissipative flux autocorrelation function evaluated in a NESS with an applied field $F_{e.}$

$$\tilde{L}'_J(s;F_e) \equiv d\tilde{L}_J(s;F_e) \big/ ds \quad . \tag{26}$$

The factor $\beta_0$ is included in the GK transform to make the expression consistent with the Green-Kubo expression for the transport coefficient at zero applied field.

Combining (22) and (24), shows that *if* the t-averaged dissipative fluxes are Gaussian, then the *nonlinear* phenomenological transport coefficient, $L(F_e)$, is given by the zero frequency Green-Kubo transform of the dissipative flux,

$$L(F_e) = \tilde{L}_J(0;F_e) = \beta_0 V \int_0^\infty dt \left\langle (J(0) - \langle J \rangle_{F_e})(J(t) - \langle J \rangle_{F_e}) \right\rangle_{F_e} . \tag{27}$$

In the zero field limit (27) reduces to the correct well known Green-Kubo expression for the linear transport coefficient, L(0). The relationship between the FT and GK expressions in the linear regime has been considered previously [4, 10, 20-22]. In the present paper, simulations are carried out to test these relationships in the nonlinear, large field regime. These numerical calculations show that this generalised Green-Kubo relation (27), is not valid, forcing us to conclude that the distribution is not sufficiently Gaussian far from equilibrium and far from the mean.

In the isoenergetic case, *if* the distribution is Gaussian, we have,

$$\lim_{(t\to\infty)} \frac{1}{t} \ln\left( \frac{p(\overline{\beta J}(t)=B)}{p(\overline{\beta J}_(t)=B)} \right) = \lim_{(t\to\infty)} \frac{2B\langle \beta J\rangle_{F_e}}{t\sigma^2_{\overline{\beta J(t)}}} = \lim_{(t\to\infty)} \frac{-2B\langle \beta J\rangle_{F_e}}{t\sigma^2_{\overline{\beta J(t)}}}. \tag{28}$$

Combining this equation with (20) shows that if the distribution is Gaussian there there must again be a trivial relation between the variance of the distribution of averaged fluxes and the nonlinear transport coefficient,

$$\langle \beta\rangle_{F_e} L(F_e) \equiv \frac{-\langle \beta J\rangle_{F_e}}{F_e} = \lim_{(t\to\infty)} \tfrac{1}{2} V t \sigma^2_{\overline{\beta J(t)}} \quad . \tag{29}$$

Were such a relation to be true at large fields it would constitute a generalised Einstein relation for the field dependent transport coefficient $L(F_e)$. In the long time limit the variance of the steady state distribution of t-averaged fluxes,

$$\sigma^2_{\overline{\beta J(t)}}(F_e) = \left\langle (\overline{\beta J}(t) - \langle \beta J\rangle_{F_e})^2 \right\rangle_{F_e}, \tag{30}$$

satisfies the equation

$$\begin{aligned} \lim_{(t\to\infty)} t\sigma^2_{\overline{\beta J(t)}}(F_e) &= 2\langle \beta\rangle_{F_e} \tilde{L}_J(0;F_e) \big/ V + \lim_{(t\to\infty)} 2\langle \beta\rangle_{F_e} \tilde{L}'_J(0;F_e) \big/ Vt \\ &= 2\langle \beta\rangle_{F_e} \tilde{L}_J(0;F_e) \big/ V \end{aligned} \tag{31}$$

where

$$\langle \beta\rangle_{F_e} \tilde{L}_J(s;F_e) \equiv V \int_0^\infty dt\, e^{-st} \left\langle (\beta J(0) - \langle \beta J\rangle_{F_e})(\beta J(t) - \langle \beta J\rangle_{F_e}) \right\rangle. \tag{32}$$

Combining (28) and (30), shows that if the distribution of the t-averaged dissipative fluxes is Gaussian, then the *nonlinear* phenomenological transport coefficient, $L(F_e)$, is given by the zero frequency Green-Kubo transform of the dissipative flux,

$$L(F_e) = \tilde{L}_J(0;F_e) \equiv V\langle\beta\rangle_{F_e}^{-1}\int_0^\infty dt\,\left\langle(\beta J(0) - \langle\beta J\rangle_{F_e})(\beta J(t) - \langle\beta J\rangle_{F_e})\right\rangle. \tag{33}$$

Not surprisingly, results of numerical tests of this relationship indicate that (33) is also not correct.

## V. NUMERICAL RESULTS

Steady state NEMD simulations of a fluid undergoing shear flow were used to test the accuracy of the expressions derived above. All simulations were carried out in two Cartesian dimensions with interactions between particles given by the Weeks-Chandler-Anderson repulsive pair potential. Note that Lennard-Jones reduced units are used in the figures and throughout this section. In both cases, simulations were carried out for systems of 200 particles and for the isokinetic system, the temperature was constrained at T = 1.0, whereas for the isoenergetic system the internal energy was constrained at E/N = 1.56032. For the isokinetic fluid, two densities, n=N/V, were considered: n = 0.4 and n = 0.8; and for the isoenergetic fluid the density was set to n = 0.8.

The SLLOD equations of motion with Lees-Edwards periodic boundary conditions were employed to model the shear flow, and a Gaussian thermostat or ergostat used to maintain a steady state [16]. The adiabatic SLLOD equations give an exact representation of shear flow arbitrarily far from equilibrium and Lees-Edwards periodic boundary conditions give the unique generalisation of periodic boundary conditions to planar Couette flow. The SLLOD equations (analogous to equation (1)) are given by:

$$\begin{aligned} \dot{\mathbf{q}}_i &= \mathbf{p}_i + \mathbf{i}\gamma y_i \\ \dot{\mathbf{p}}_i &= \mathbf{F}_i - \mathbf{i}\gamma p_{yi} - \alpha \mathbf{p}_i \end{aligned} \qquad (34)$$

where $\gamma$ is the strain rate and $\alpha$ is the isokinetic or isoenergetic thermostat multiplier. When the kinetic energy is a constant of motion,

$$\alpha_K = \frac{\sum_{i=1}^{N} \mathbf{F}_i \cdot \mathbf{p}_i - \gamma p_{xi} p_{yi}}{\sum_{i=1}^{N} \mathbf{p}_i \cdot \mathbf{p}_i} \qquad (35)$$

while if the internal energy is a constant of motion,

$$\alpha_E = \frac{-\gamma P_{xy} V}{\sum_{i=1}^{N} \mathbf{p}_i \cdot \mathbf{p}_i} \tag{36}$$

where $P_{xy}$ is the xy element of the pressure tensor,

$$P_{xy} V = \sum_{i=1}^{N} p_{xi} p_{yi} - \frac{1}{2} \sum_{i,j=1}^{N} x_{ij} F_{yij}, \tag{37}$$

which is the dissipative flux: $J \equiv P_{xy}$. The nonlinear shear viscosity, $\eta(\gamma)$ is the nonlinear transport coefficient calculated using this algorithm. We note that in contrast to the discussion above, the dissipative flux for shear flow is even under the time reversal mapping, $M^T(\mathbf{x}, \mathbf{y}, \mathbf{p}_x, \mathbf{p}_y) = (\mathbf{x}, \mathbf{y}, -\mathbf{p}_x, -\mathbf{p}_y)$ and the strain rate is odd. However we can choose the strain rate to be even and the dissipative flux odd by employing the Kawasaki mapping [16], $M^K(\mathbf{x}, \mathbf{y}, \mathbf{p}_x, \mathbf{p}_y) = (\mathbf{x}, -\mathbf{y}, -\mathbf{p}_x, \mathbf{p}_y)$.

Firstly we carried out simulations to show that in the long time limit, for an isokinetic system the variance of the distribution of $\{\bar{J}(t)\}$ is related to the zero frequency Green-Kubo transform of $\bar{J}(t)$ by equation (24), and for an isoenergetic system, the variance of the distribution of $\{\overline{J\beta}(t)\}$ is related to the zero frequency Green-Kubo transform of $\overline{J\beta}(t)$ by equation (31). The behaviour at various strain rates was examined and the results are shown in Figure 3. Equations (24) and (31) are found to be verified in all cases.

We tested the nonlinear Green-Kubo relations (27) and (33) for the systems described above and the results are shown in Figure 4. Since $\tilde{L}(F_e)$ is also related to the variance of the distribution of $\{\bar{J}(t)\}$ or $\{\overline{J\beta}(t)\}$ as $t \to \infty$ for the isokinetic or isoenergetic system, respectively, results obtained directly from the variance are also presented. Clearly the equivalence of $L(F_e)$ and $\tilde{L}(F_e)$ is only observed at small fields. At intermediate fields the Green-Kubo transform of the dissipative flux $\tilde{L}(F_e)$, underestimates the actual transport coefficient while at high fields $\tilde{L}(F_e)$, overestimates the transport coefficient. We conclude that nonlinear Green-Kubo relations (27, 33) are **not** valid in the far from equilibrium regime.

## VI. DISCUSSION

In the zero field limit, thermostatted linear response theory can be used to determine the field dependent transport coefficients. For the isokinetic response:

$$\lim_{F_e \to 0} L(F_e) = -\lim_{F_e \to 0} \frac{\langle J \rangle_{F_e,K}}{F_e} = \beta_0 V \int_0^{\infty} dt \langle J(0)J(t) \rangle_{0,K} \tag{38}$$

where the ensemble average $\langle \ \rangle_{0,K}$ is over the equilibrium isokinetic ensemble. This expression derived from linear response theory is identical to (27) in the limit $F_e \to 0$, which was derived using both the CLT and the FT. The results in Figure 4 confirm the agreement of (38) and (27) in the zero field limit with linear response theory.

This work also shows that in the zero field limit, one can calculate linear transport coefficients by considering the limiting long time variance, $\sigma^2_{\bar{J}(t)}(F_e = 0)$, of the distributions of $\bar{J}(t)$ (24, 31), rather than by computing autocorrelation functions of the dissipative flux and then performing the appropriate long time integrals. The variance of the t-averaged flux therefore provides an alternative route to the *linear* transport coefficients and equations (24, 31) thus provide useful Einstein routes to linear transport coefficients.

We now turn to the question of why the nonlinear Green-Kubo and Einstein expressions fail, far from equilibrium. A necessary condition for the CLT is the statistical independence of the sample averages. A breakdown of this independence could be responsible for the breakdown of the CLT, Einstein and Green-Kubo expressions. However, trajectory segments that are much longer than the Maxwell time, $\tau_M$, which characterises the decay of the autocorrelation function of the dissipative flux autocorrelation function, should have *no* correlations between successive samples of $\bar{J}(t)$. Thus, regardless of the distribution of J(t) we expect that for long enough t, the CLT will apply. Figure 5 compares the decay time autocorrelation function of J(t) for different applied fields. At moderate fields $\tau_M$ is *less* than it is at equilibrium. Only at very large fields does $\tau_M$ increase. This means that possible decay time divergences or anomalies are ***not*** responsible for the breakdown of the nonlinear

Einstein and GK expressions (33, 27). Further, if one computes the distribution of $\bar{J}(t)$, for various values of t, one cannot *observe* departures from Gaussian behaviour for values of t >> $\tau_M$, in the neighbourhood of the mean current.

Figure 6 compares the distributions of J(t) (the distribution of the instantaneous flux) and $\bar{J}(t)$ for an equilibrium and nonequilibrium system with a strain rate which ensures it is in the nonlinear regime (T=1.0, n=0.8, $\gamma$=1.0). While the skewness, $\gamma_1$, and kurtosis, $\kappa$, for the instantaneous equilibrium J(t) distribution are zero within error bars which is consistent with a Gaussian distribution, the skewness is non-zero for the sheared system ($\gamma_1 = -0.23\pm0.01$, $\kappa = 0.13\pm0.04$, respectively). The distributions of the time-averaged fluxes, $\bar{J}(t)$, were obtained for a trajectory segment of length t = 4.0 and both distributions, as expected, *appear* Gaussian. The skewness of the distribution for the sheared system is $\gamma_1 = -0.064\pm0.004$, and the kurtosis $\kappa = -0.02\pm0.02$. Thus *on the basis of these tests* although for a sheared system the distribution of J(t) is not Gaussian the distribution of $\bar{J}(t)$, for a trajectory segment of length t = 4.0, is on the scales shown in Figure 6, already indistinguishable from a Gaussian.

As noted in references [21, 22], the distribution of $\bar{J}(t)$ and $\overline{J\beta}(t)$ cannot be exactly Gaussian because the values of these variables are bounded. In practice however these bounds are so large that they become irrelevant in the limit $t \to \infty$ where the t-averaged distributions collapse to zero variance distributions. Moreover, the bounds still apply in the zero field limit where the Green-Kubo and Einstein expressions are all valid. Thus the boundedness of the fluxes cannot be the responsible for the breakdown of the nonlinear GK and Einstein expressions.

If we examine the derivation of equations (27) and (33) more closely, it can be seen that in order to obtain a GK expression we require the distribution at ***both*** $\bar{J}(t) = \bar{J}_+(t)$ and $\bar{J}(t) = -\bar{J}_-(t)$ be well approximated by a Gaussian for times sufficiently long that the GK integrals have converged, t >> $\tau_M(F_e)$ (see Appendix for details) [13, 14]. Any deviations from the behaviour indicated by (21) and (28) will be related to the relative deviation of the distribution from a Gaussian at both $\bar{J}_+(t)$ and $\bar{J}_-(t)$. It is therefore of interest to consider the rate of convergence to a Gaussian. The magnitude of the relative deviation of the distribution $p((\overline{J}(t)-\overline{\overline{J}})/\sigma_{\overline{J}(\tau_M)})$ from a normalised Gaussian generally increases with the separation of $\overline{J}(t)$ from the mean $\overline{\overline{J}}$ for sufficiently large separations (see, for example, section 7.2 of [13]). Here $\sigma_{\overline{J}(\tau_M)}$ is the standard deviation of the distribution $p(\overline{J}(t))$ when $t = \tau_M$. In the $t \to \infty$ limit, at fixed $\overline{J}(t)$, the magnitude of the relative devaition of $p(\overline{J}(t))$ from a Gaussian becomes infinite. This means that in the $t \to \infty$ limit, the CLT gives information which is not sufficiently precise to derive Green-Kubo relations for non-zero applied fields.

We illustrate this point in more detail. Suppose that $\overline{J}(t) = \overline{J}_+(t)$ is equal to the mean current, $\overline{\overline{J}}$; then the conjugate trajectories will have $\overline{J}(t) = \overline{J}_-(t) = -\overline{\overline{J}}$. Clearly $|\overline{J}_-(t) - \overline{\overline{J}}| = 2L(F_e)F_e$. Therefore using equation (45) of Appendix A, we find in the $t \to \infty$ limit, *except* when $F_e = 0$,

$$\left|\overline{J}_-(t) - \overline{\overline{J}}\right| / \sigma_{\overline{J}(\tau_M)} = 2L(F_e)F_e / \sigma_{\overline{J}(\tau_M)} \approx F_e\sqrt{2VtL(F_e)\beta} \to \infty . \tag{39}$$

For any non zero field, if $\overline{J}(t) = \overline{\overline{J}}$, then as t increases, the value of $\overline{J}_-(t)$ moves further and further into the wings of the normalised distribution where the magnitude of the relative deviation of $p(\overline{J}(t))$ from a Gaussian grows without bound. Strictly speaking therefore, in the infinite time limit, for any finite field, the relative deviation of $p(\overline{J}(t))$ from a Gaussian, evaluated in the neighbourhood of the mean anticurrent, $-\overline{\overline{J}}$ grows without bound and nonlinear Green-Kubo relations cannot be derived. However, in practice one does not need to take the infinite time limit. Considering the shift in the mean value of the dissipative flux with field shows that the nonlinear GK expression will be *approximately* correct provided,

$$F_e \le O\left(1/\sqrt{\beta V_M \tau_M(F_e)L(F_e)}\right), \tag{40}$$

where $V_M$ is the minimum volume required for transport coefficient to be approximately equal to its large system, limiting value [24]. Clearly the nonlinear GK relations satisfy this relation only in a small neighbourhood *including* $F_e = 0$. For the systems studied here, equation (40) predicts that the nonlinear GK relations will be *approximately* correct provided $\gamma <\sim 10^{-1}$. This is in agreement with experimental observations given in Figures 4(a),(b), 4(c).

**Acknowledgements**

We would like to thank the Australian Research Council for the support of this project. The helpful discussions and comments from Professor E.G.D. Cohen, Professor G. Gallavotti, Dr C. Jarzynski and Dr R. van Zon are also gratefully acknowledged. DJE would like to thank the National Institute of Standards and Technology, Boulder, Colorado for support.

**Appendix.**

The variance of the time-averaged dissipative flux is given by,

$$\begin{aligned}\sigma^2_{\bar{J}(t)} &= \left\langle (\bar{J}(t) - \langle J \rangle)^2 \right\rangle \\ &= \left\langle \frac{1}{t^2}\left(\int_0^t ds_1 \Delta J(s_1)\right)\left(\int_0^t ds_2 \Delta J(s_2)\right)\right\rangle \\ &= \frac{1}{t^2}\int_0^t ds_1 \int_0^t ds_2 \left\langle \Delta J(s_1)\Delta J(s_2)\right\rangle\end{aligned} \tag{40}$$

where $\Delta J(t) = \bar{J}(t) - \langle J \rangle$. Using a change of variables: $\tau_1 = s_1 - s_2$ and $\tau_2 = s_1 + s_2$ this integral can be written:

$$\begin{aligned}\sigma^2_{\bar{J}(t)} &= \frac{1}{2t^2}\int_0^t d\tau_2 \int_{-\tau_2}^{\tau_2} ds_2 \left\langle \Delta J(\tfrac{1}{2}(\tau_1 + \tau_2))\Delta J(\tfrac{1}{2}(\tau_2 - \tau_1))\right\rangle \\ &+ \frac{1}{2t^2}\int_0^t d\tau_2 \int_{-\tau_2}^{\tau_2} ds_2 \left\langle \Delta J(\tfrac{1}{2}(\tau_1 + \tau_2 - \tau))\Delta J(\tfrac{1}{2}(\tau_2 - \tau_1 - \tau))\right\rangle\end{aligned} \tag{41}$$

Since correlation functions are invariant under a time translation in the steady state, and using the symmetry of the functions we obtain,

$$\sigma^2_{\bar{J}(t)} = \frac{2}{t^2}\int_0^t d\tau_2 \int_0^t d\tau_1 \left\langle \Delta J(\tau_1)\Delta J(0)\right\rangle. \tag{42}$$

Changing the order of integration gives:

$$\begin{aligned}\sigma^2_{\bar{J}(t)} &= \frac{2}{t^2}\int_0^t d\tau_1 \int_{\tau_1}^t d\tau_2 \left\langle \Delta J(\tau_1)\Delta J(0)\right\rangle \\ &= \frac{2}{t^2}\int_0^t d\tau_1 \int_{\tau_1}^t d\tau_2 \left\langle \Delta J(\tau_1)\Delta J(0)\right\rangle \ . \\ &= \frac{2}{t^2}\int_0^t d\tau_1 \left\langle \Delta J(\tau_1)\Delta J(0)(t - \tau_1)\right\rangle\end{aligned} \tag{43}$$

Therefore, for any steady state system, at all times:

$$t\sigma^2_{\bar{J}(t)}(F_e) = 2\int_0^t ds \left\langle (J(0) - \langle J \rangle_{F_e})(J(s) - \langle J \rangle_{F_e})\right\rangle - \frac{2}{t}\int_0^t ds \left\langle (J(0) - \langle J \rangle_{F_e})(J(s) - \langle J \rangle_{F_e})\right\rangle s. \tag{44}$$

At any time greater than the time required for the time correlation function to decay to zero, $t_C > t_M$, in an isokinetic system, $\int_0^{t_C} ds \left\langle (J(0) - \langle J \rangle_{F_e})(J(s) - \langle J \rangle_{F_e})\right\rangle = \tilde{L}_J(0;F_e)/(\beta_0 V)$ and $\int_0^{t_C} ds \left\langle (J(0) - \langle J \rangle_{F_e})(J(s) - \langle J \rangle_{F_e})\right\rangle s = -\tilde{L}'_J(s;F_e)/(\beta_0 V)$. Therefore,

$$t_C \sigma^2_{\bar{J}(t_C)}(F_e) = \frac{2\tilde{L}_J(0;F_e)}{\beta_0 V} + \frac{2\tilde{L}'_J(0;F_e)}{\beta_0 V t_C}. \tag{45}$$

*If* the distribution is Gaussian at $\bar{J}(t)$ and $-\bar{J}(t)$ at $t_C$, then assuming that the second term of (45) is negligible and that the FT is true at $t = t_C$, combining (20) and (45) gives,

$$\frac{-\langle J \rangle_{F_e}}{F_e} = \tfrac{1}{2}\beta_0 V t_C \sigma^2_{\bar{J}(t_C)} = \tilde{L}_J(0;F_e). \tag{46}$$

That is, a GK expression is valid.

**Figure Captions**

**Figure 1.** The shear stress, $P_{xy}$ for trajectory segments from a simulation of 200 disks at T = $\Sigma p_i^2/2mNk_B$ = 1.0, and n = N/V = 0.4. The trajectory segment, $\boldsymbol{\Gamma}_{(1,3)}$, was obtained from a forward time simulation. At t = 2, a time reversal map was applied to $\boldsymbol{\Gamma}_{(2)}$, to give $\boldsymbol{\Gamma}_{(5)}$ (for the SLLOD equations of motion the time reversal map is the Kawasaki map (x, y, $p_x$, $p_z$, $\gamma$)$\rightarrow$ (x, -y, -$p_x$, $p_z$, $\gamma$)). Forward and reverse time simulations from this point give the trajectory segments $\boldsymbol{\Gamma}_{(5,6)}$ and $\boldsymbol{\Gamma}_{(5,4)}$, respectively. If one inverts $P_{xy}$ in $P_{xy}$ = 0 and inverts time about t = 2, one transforms the $P_{xy}(t)$ values for the anti-segment $\boldsymbol{\Gamma}_{(4,6)}$ into those for the conjugate segment, $\boldsymbol{\Gamma}_{(1,3)}$.

**Figure 2.** Schematic diagram showing two disconnected subvolumes of phase space $\delta V_1(0)$, $\delta V_1'(0)$ from which trajectories originate which, after a time $2\tau$, have time averaged values of $\alpha$ which lie in the range: $(\overline{\alpha}_+(2\tau), \overline{\alpha}_+(2\tau) + d\alpha)$. At time $2\tau$ these volumes evolve to: $\delta V_1(2\tau)$ = $\delta V_3$, $\delta V_1'(2\tau) = \delta V_3'$. See equation (16).

**Figure 3.** Calculation of $\tilde{L}_J(0;Fe)$ from the variance of distributions of the t-averaged dissipative flux for simulations at: (a) constant temperature with T = 1.0, n = 0.4, $F_e$ = 0.0 (unfilled circles), $F_e$ = 1.0 (filled circles) and $F_e$ = 2.0 (squares); (b) constant temperature with T = 1.0, n = 0.8; $F_e$ = 0.0 (unfilled circles), $F_e$ = 0.5 (filled cirlces) and $F_e$ = 1.0 (squares); and (c) constant internal energy with E/N = 1.56032, n = 0.8, $F_e$ = 0.0 (unfilled circles), $F_e$ = 0.3 (triangles) and 0.5 (filled circles). The crosses show $\tilde{L}_J(0;Fe)$ determined from the zero frequency Green-Kubo transform of the dissipative flux.

**Figure 4.** The filled circles show the viscosity as a function of strain rate for systems at (a) constant temperature with T = 1.0 and n = 0.4; (b) constant temperature with T = 1.0 and n = 0.8; and (c) constant internal energy with E/N = 1.56032 and n = 0.8. The crosses are predictions determined from the Green-Kubo expression (equation (27) for the isokinetic case and equation (33) for the isoenergetic case) and the squares are predictions from the variance

(equation (22) for the isokinetic case and equation (29) for the isoenergetic case).

**Figure 5.** The decay of the shear stress time correlation function, $\int_0^t ds \left\langle P_{xy}(0)P_{xy}(s)\right\rangle$, at equilibrium and in nonequilibrium steady states for a systems at constant temperature with T = 1.0 and n = 0.4. The full line is at equilibrium ($\gamma = 0$) the dotted line is for a simulation with a strain rate of $\gamma = 1$ and dashed line with a strain rate of $\gamma = 2$.

**Figure 6.** (a) The instantaneous (small circles) and time-averaged distribution (squares, t = 4.0) of the dissipative flux for an equilibrium system at T = 1.0 and n = 0.8. (b) The instantaneous (small circles) and time-averaged distribution (squares, t = 4.0) of the dissipative flux for a nonequilibrium system at T = 1.0, n = 0.8 and with $\gamma = 1.0$.

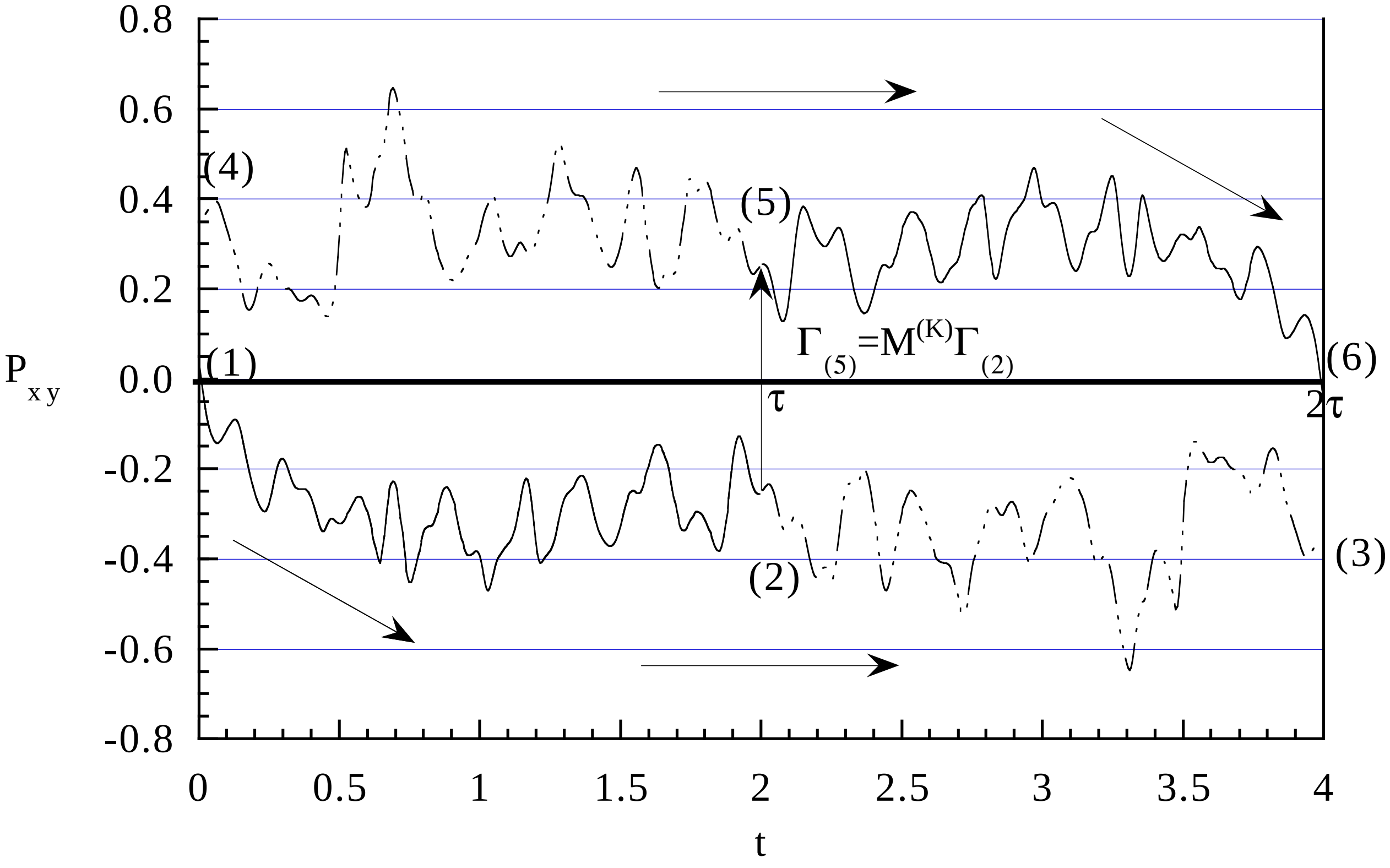

Figure 1
Searles and Evans

$$\delta V_3 / \delta V_1 = \delta V_{3'} / \delta V_{1'} = \exp[-6N\tau\bar{\alpha}_+(2\tau)]$$

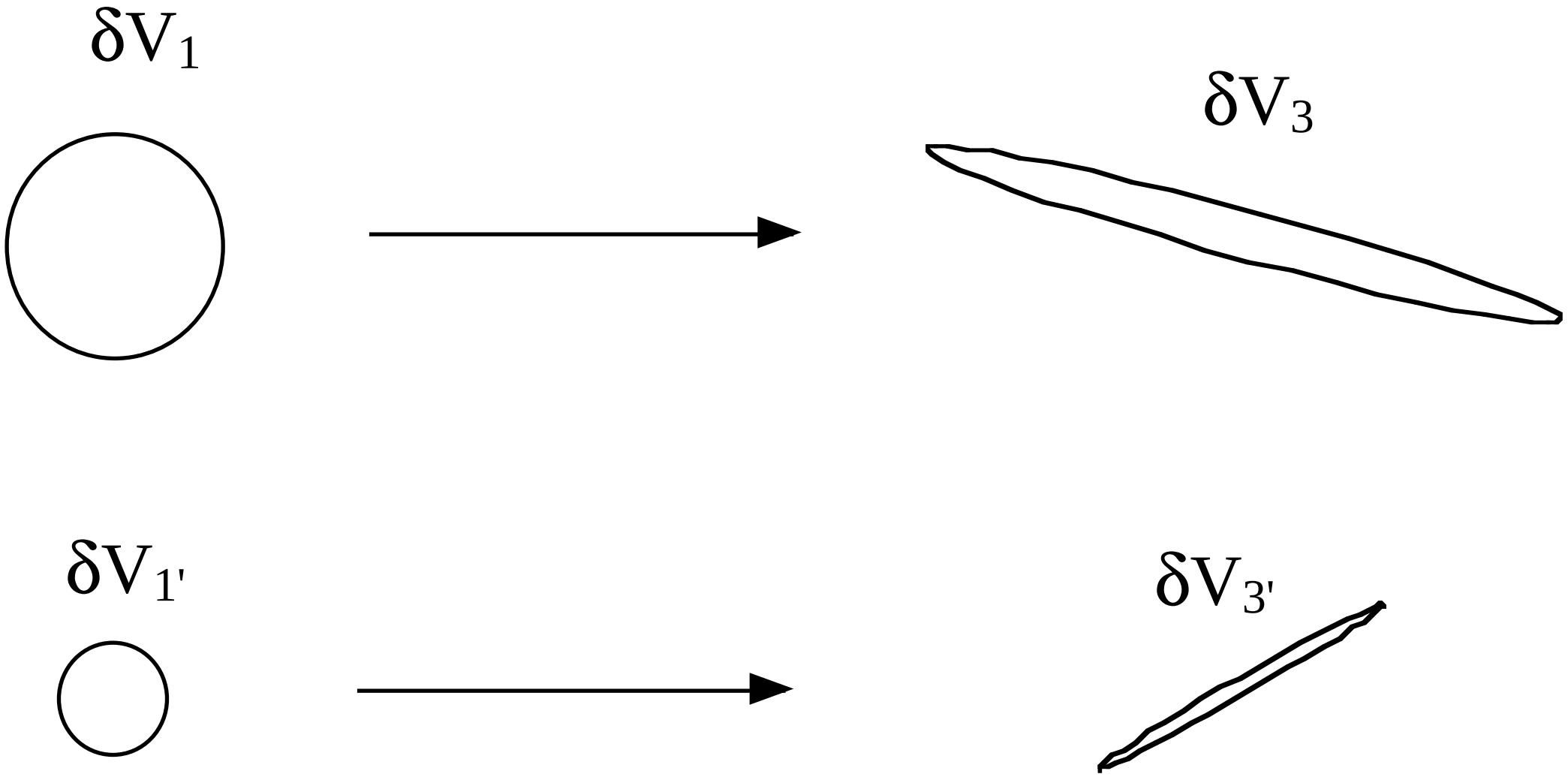

$$\bar{\alpha}_+(2\tau) = \frac{1}{2\tau}\int_0^{2\tau} dt\, \alpha(t;\mathbf{\Gamma}_1) = \frac{1}{2\tau}\int_0^{2\tau} dt\, \alpha(t;\mathbf{\Gamma}_{1'})$$

Figure 2
Searles and Evans

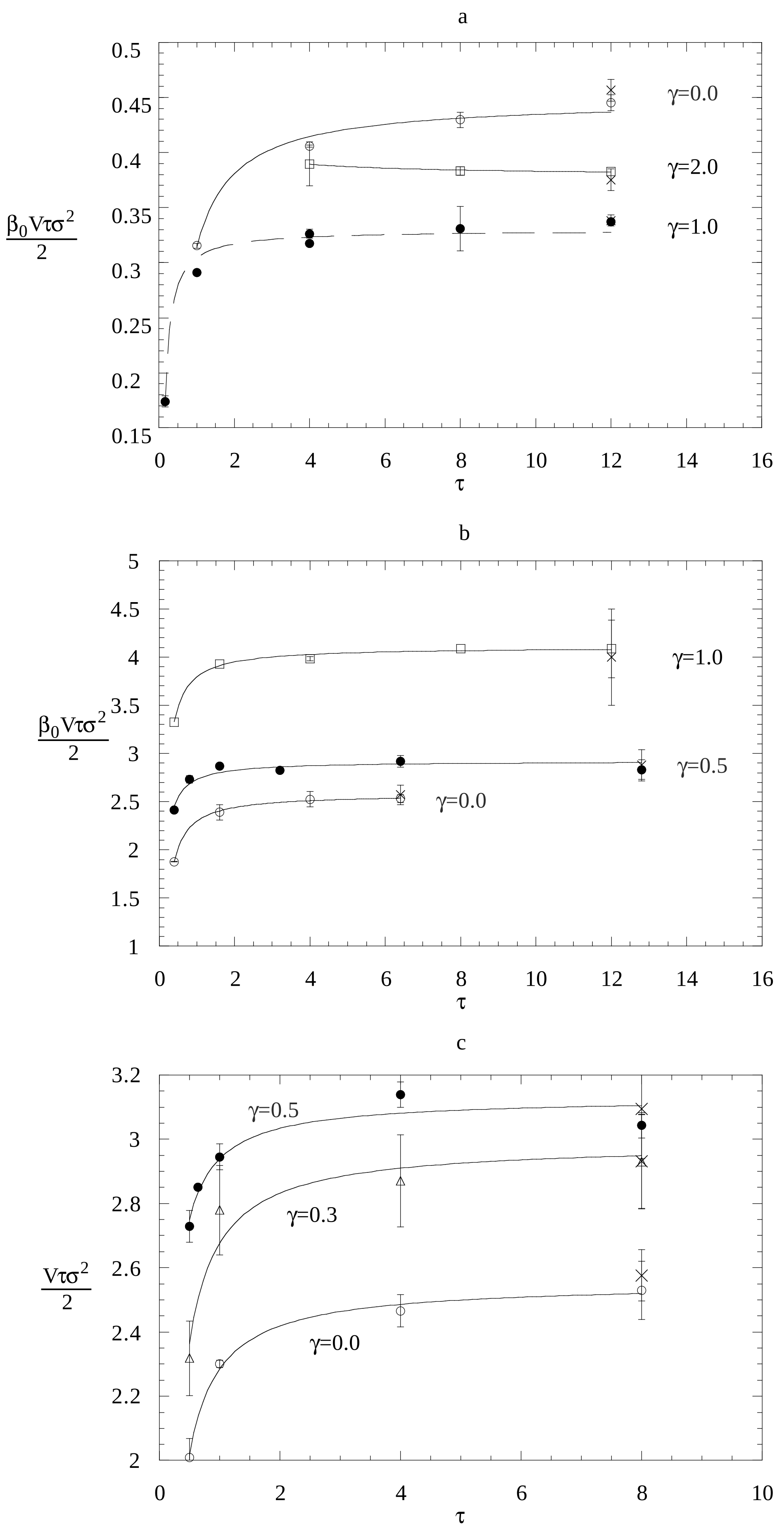

Figure 3
Searles and Evans

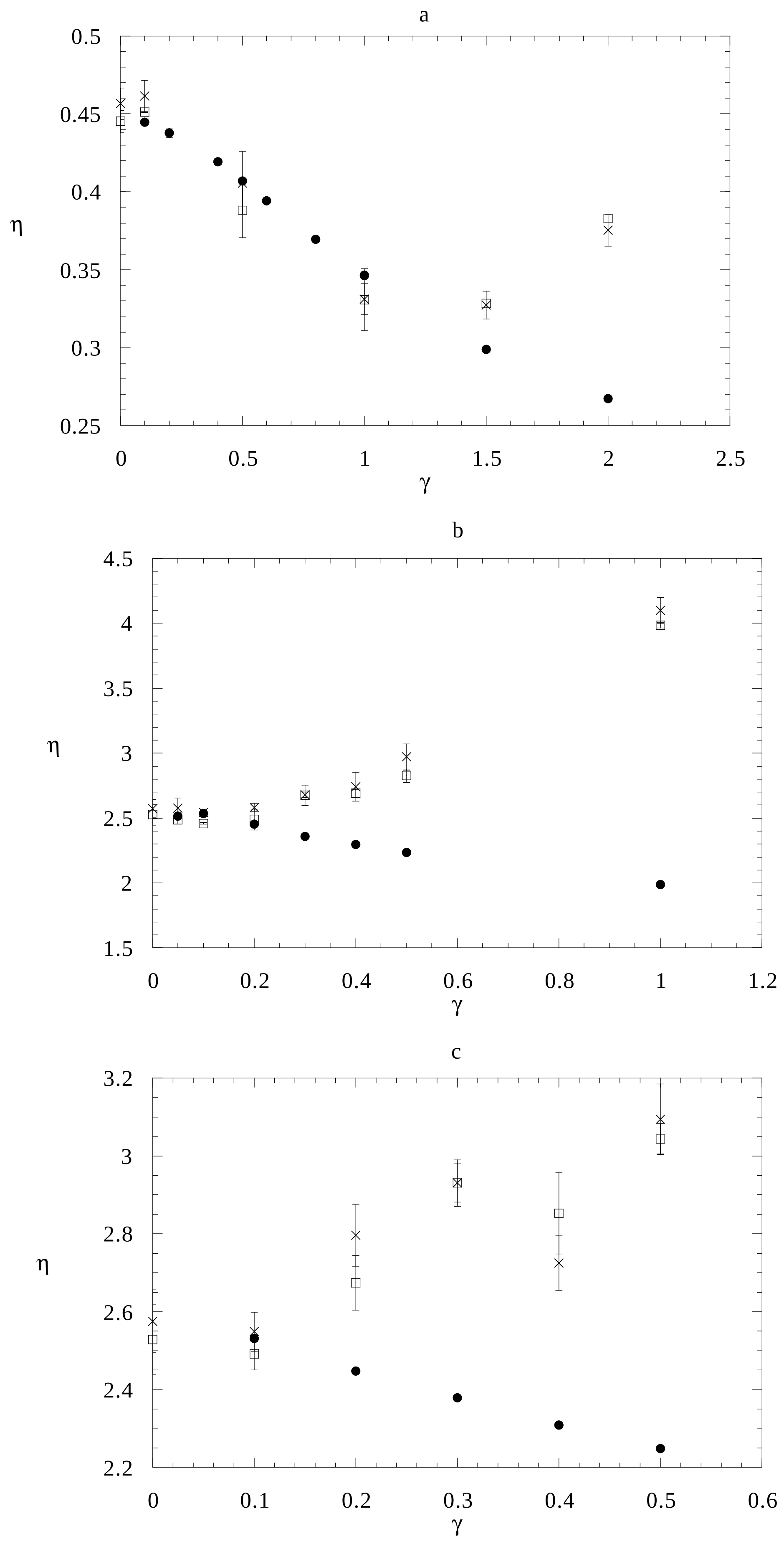

Figure 4
Searles and Evans

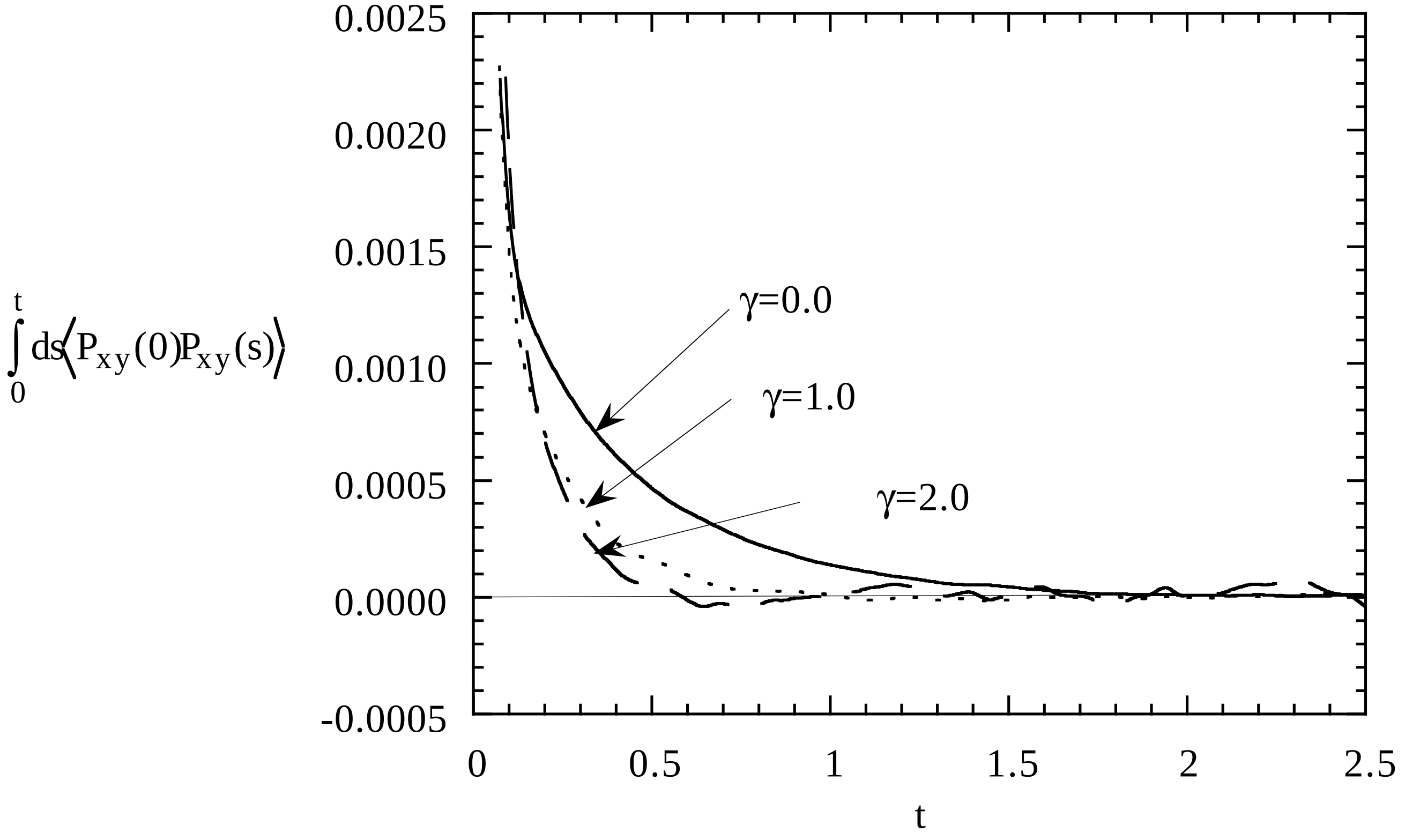

Figure 5
Searles and Evans

a

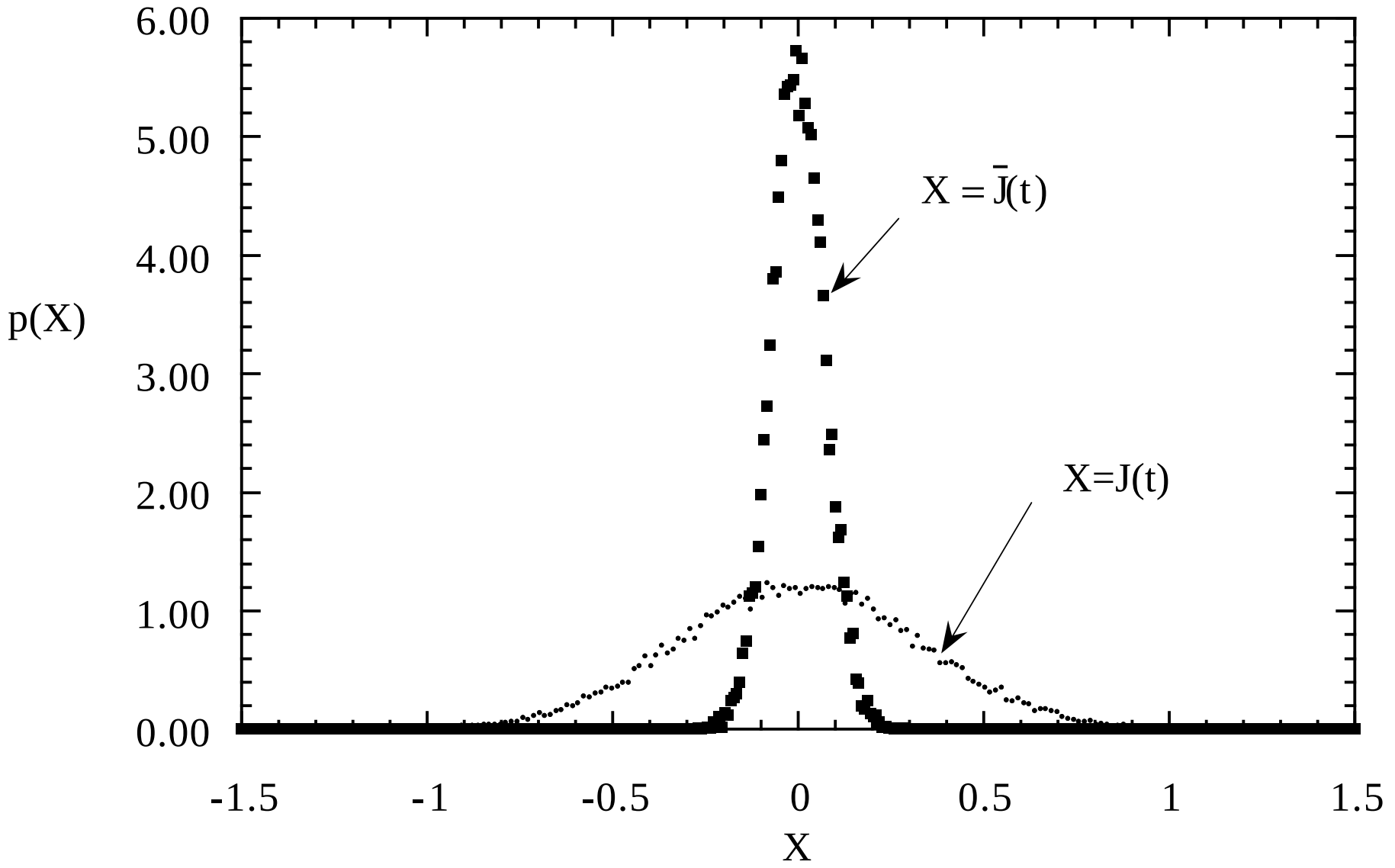

b

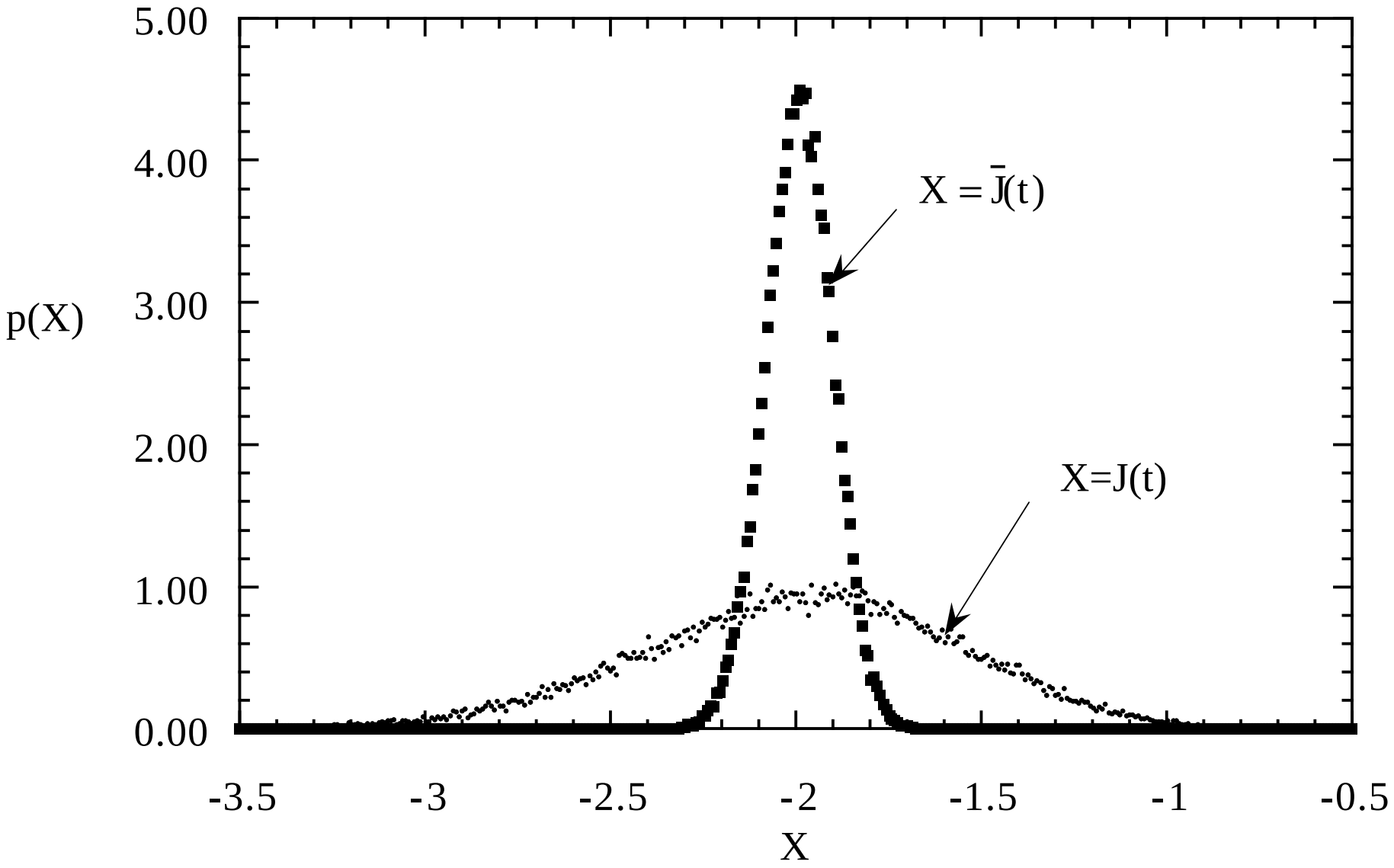

Figure 6
Searles and Evans